\documentclass[aps,pra,a4paper, amsfonts, amssymb, amsmath, reprint, showkeys, nofootinbib, twoside]{revtex4-1}

\usepackage[english]{babel}
\usepackage[utf8]{inputenc}
\usepackage[colorinlistoftodos, color=green!40, prependcaption]{todonotes}
\usepackage{graphicx}
\usepackage{bm}

\usepackage{xcolor}

\begin{document}

\title{Longitudinal Angular Momentum in Mie Scattering {from a Magneto-optically Active Sphere}: \\ QED Correction to the Einstein-De Haas effect }

\author{B.A. van Tiggelen}
\email[]{Bart.Van-Tiggelen@lpmmc.cnrs.fr}
\affiliation{Univ. Grenoble Alpes, CNRS, LPMMC, 38000 Grenoble, France}

\date{\today} 

\begin{abstract}
This work investigates the angular momentum  {induced by } electromagnetic quantum  {fluctuations} in a dielectric Mie sphere.  {{The magneto-optical activity} creates longitudinal electric modes with broken symmetry of magnetic quantum numbers}. When excited by electromagnetic quantum  {fluctuations},  angular momentum is created
by a vortex of the  Poynting vector that decays as $1/r$ from the center of the sphere.  This longitudinal angular
momentum,  connected to the vector potential $\mathbf{A}(\mathbf{r},t)$, emerges as a QED modification
  {of the traditional diamagnetic }
Einstein-De Haas effect in which an applied external magnetic field - via its action on microscopic magnetism - enforces macroscopic rotation.
\end{abstract}


\maketitle

\section{Introduction}

Electromagnetic{ fields} possess angular momentum. It often emerges as spin or orbital motion \cite{laguerre, Bliot}, associated  with transverse
electric and magnetic fields.
''Longitudinal" angular momentum is associated
with longitudinal electric waves for which $\bm{\nabla}\cdot \mathbf{E}\neq 0$ at some points in space but  $\bm{\nabla}\times \mathbf{E}=0$ everywhere.
{Microscopically they are created by charges, {either moving or not}.}
A classical example is when an {immobile} electrical charge and a magnetic dipole are brought close together.
The longitudinal Coulomb field of the charge and the magnetic dipole field
induce together a Poynting vector field $c_0 \mathbf{E} \times \mathbf{B}/4\pi $  {that circulates around the magnetic moment of the dipole and } with finite angular momentum \cite{feynman}. In this work we describe how
electromagnetic quantum  {fluctuations }induce
longitudinal angular momentum inside {a magneto-optically active}  ``Mie" sphere.  {This is accompanied by a topology of the Poynting vector similar to the one in the example above.}
For a magneto-optically active Mie sphere, longitudinal electric fields {do} not only exist
on its boundary, but also inside the sphere.

More generally, the interaction of   electromagnetic quantum fluctuations  with matter creates both ``Casimir"  energy  \cite{milonni,milton} and
``Casimir"  momentum \cite{feigel}. For macroscopic objects both
suffer often from power-law divergencies at high photon energies. {Some divergencies in Casimir energy for instance,  are not observable and removed e.g.} by dimensional regularization \cite{milton},
others require QED mass renormalization \cite{kawka,manuel}.
The classical electromagnetic ``Abraham" momentum was observed first by Walker etal. \cite{Walker} and recently its predicted  {QED} corrections have been
searched for \cite{geert}.

Longitudinal  angular momentum  {has a direct connection to the electromagnetic vector potential} $\mathbf{A}(\mathbf{r},t)$. A simple example where it emerges is a
charged particle (an electron {with charge $e <0$ }) {with no spin} and mass $m$ rotating
 around a homogeneous, time-dependent
magnetic field. For a time-dependent magnetic field {$\mathbf{B}_0$ with constant orientation along the $z$-direction},
Faraday's induction law \cite{jackson} {implies the existence of}  a transverse electric field $\mathbf{E} = (\mathbf{r}\times d\mathbf{B}_0/dt) /2c_0$
along the orbit of the rotating charge. The total Lorentz force on the charge is ({in Gaussian units})
\begin{equation*}
   \mathbf{ F} = m \frac{d\mathbf{v}}{dt} = \frac{1}{2}\frac{e }{c_0}\mathbf{r} \times \frac{d\mathbf{B}_0}{dt} + \frac{e}{c_0}\frac{d\mathbf{r}}{dt} \times \mathbf{B}_0
\end{equation*}
 {If both the velocity} $\mathbf{v}$  {and the position vector }$\mathbf{r}$  {of the charge are located in the plane perpendicular to the magnetic field, the torque on the charge  is}
   \begin{equation*}
  \mathbf{ r}\times \mathbf{ F} =  \frac{d\mathbf{L}_{\mathrm{kin}}}{dt} = \frac{e}{2c_0} \frac{d}{dt} \left[ \mathbf{r} \times (\mathbf{r} \times \mathbf{B}_0 ) \right]
\end{equation*}
 {This expression is equivalent to the } Lenz law stating that the sum of kinetic angular momentum $ \mathbf{L}_{\mathrm{kin}} = (\mathbf{r }\times m\mathbf{v})_z$
 along the direction $z$  and enclosed magnetic
flux $\Phi = B_0A$ is conserved in time,
\begin{equation}\label{lenz}
    \frac{d}{dt } \left[ L_{\mathrm{kin},z} + \frac{e}{2\pi c_0} \Phi \right] = 0
\end{equation}
 {The link to  electromagnetic angular momentum can be established by considering} the longitudinal Coulomb field
 $\mathbf{E}_L(\mathbf{r}') = -\bm{\nabla}' \phi $ of the moving charge at $\mathbf{r}(t)$ and
the vector potential $\mathbf{A}_0(\mathbf{r}',t)$ associated with the external magnetic field.
The longitudinal electromagnetic angular momentum is given by \cite{cohen}

\begin{equation*}
    \mathbf{J}_{\mathrm{long}}= \frac{1}{4\pi c_0}\int d^3 \mathbf{r}' \, \mathbf{r}' \times [\mathbf{E}_L(\mathbf{r}') \times \mathbf{B}_0(\mathbf{r}')]
\end{equation*}
 {If  we assume a  magnetic field that is homogeneous in the vicinity of the charge and zero far away,
it is straightforward to demonstrate that $\mathbf{J}_{\mathrm{long}}$ resides entirely at the position $\mathbf{r}$ of the charge.
In the Coulomb gauge for $\mathbf{A}_0(\mathbf{r},t)$ we find that  $\mathbf{J}_{\mathrm{long}}= e \mathbf{r} \times \mathbf{A}_0(\mathbf{r},t)/c_0 $.
For $ \mathbf{A }_0(\mathbf{r},t) = \mathbf{B }_0(t)\times \mathbf{r}/2$ in the
vicinity of  the charge {where the magnetic field is assumed homogeneous}, Eq.~(\ref{lenz}) becomes equivalent to the conservation of  kinetic plus longitudinal electromagnetic angular
momentum. }

 {The connection of longitudinal electromagnetic angular momentum to the vector potential becomes explicit} in the quantum-mechanical description that has Hamiltonian,
\begin{equation}\label{QMlenz}
    H = \frac{1}{2m} \left( \mathbf{p} - \frac{e}{c_0} \mathbf{A}_0(\mathbf{r},t) \right)^2
\end{equation}
Here $\mathbf{p}$ is the canonical momentum operator, i.e canonical to $\mathbf{r}$ with $[r_i,p_j]= i\hbar \delta_{ij}$.
The kinetic momentum $\mathbf{p}_{\mathrm{kin}} = \mathbf{p} - e\mathbf{A}_0/c_0$ in  Eq.~(\ref{QMlenz}) is manifestly gauge-invariant. However,
the conserved quantity of this model is the gauge-variant canonical angular momentum $J_{\mathrm{can},z} = (\mathbf{r} \times\mathbf{ p})_z$ since
it commutes with $H$  {and has no explicit time-dependence despite a possible
time-dependent $\mathbf{A}_0(\mathbf{r},t)$}. Hence, $\mathbf{J}_{\mathrm{can}} = \mathbf{J}_{\mathrm{kin}} +
\mathbf{J}_{\mathrm{long}} $ is conserved with $ \mathbf{J}_{\mathrm{long}}= e\mathbf{r}\times \mathbf{A}_0/c_0$.
When electromagnetic quantum  {fluctuations } {are added to the vector potential in Eq.~(\ref{QMlenz}), the former arguments still apply, although spin
and orbital motion associated} with the transverse
electromagnetic field  provide the dominant QED correction to total angular momentum \cite{physres}.

The conservation of canonical angular momentum remains valid in the presence of a rotationally-symmetric confining potential for the charge $e$,
and provided that
no optical transitions occur that usually involve
electromagnetic spin. This is true for  the
ground state $| \Psi_0\rangle$ of an isotropic harmonic oscillator with eigenfrequency $\omega_0$. { From now on we will
assume the external field }$\mathbf{B}_0$   {to change slow enough to consider it stationary in the calculation of the
angular momenta}.
Upon switching on the magnetic field slowly,
\begin{equation}\label{ho}
  \langle J_{\mathrm{long},z} \rangle = \frac{eB_0(t)}{2c_0} \langle \Psi_0|  x^2 + y^2|\Psi_0 \rangle =\hbar \times  \frac{\omega_c}{\omega_0}
\end{equation}
with $\omega_c = eB_0/2mc_0$ the classical cyclotron frequency. {Note that for an electron this frequency is  negative. For convenience we have added the factor of $1/2$ so that Larmor
and cyclotron frequency coincide if the Land\'e factor of magnetic moment is one}.  Because total angular momentum is conserved,
a kinetic angular momentum $-\langle J_{\mathrm{long},z} \rangle$ is
achieved by the matter,{  and thus a  magnetic moment,}

\begin{equation*}
    \mathbf{m}= \frac{e}{2mc_0}\,  {L}_{\mathrm{kin}} = -\frac{\hbar e \omega_c}{2mc_0 \omega_0}\,  \hat{\mathbf{z}}
\end{equation*}

{This is always opposite to the magnetic field direction regardless of the sign of the charge.} Since $|\omega_c| /\omega_0 \ll 1$
this diamagnetic effect is much smaller than in paramagnetism, where spins of order $\hbar$ are involved.

In this work we study the angular momentum of the quantum  {fluctuations}  that emerge from their interactions with a macroscopic,
dielectric Mie sphere \cite{newton, vdhulst} subject to a
magnetic field \cite{lacoste}. The sphere is described by a macroscopic complex-valued, frequency-dependent  dielectric tensor
assumed homogeneous for $r< a$. In a homogeneous medium {the magneto-optical activity induces the } Faraday effect that lifts the degeneracy of circular polarization in the dispersion
law, {relating frequency to wave number and resulting in different light speeds for opposite circular polarization. }
For a  Mie sphere the {magneto-optical activity} lifts the degeneracy of the magnetic quantum numbers in the eigenvalues of the {Vector }Helmholtz equation,  {akin to} the Zeeman effect in atoms.
Not less important, the {magneto-optical activity } also couples longitudinal and transverse eigenmodes of the Vector Helmholtz equation
so that longitudinal electric fields reside not only at the surface but also inside the sphere.

  {In the traditional picture of }the Einstein-De Haas effect \cite{EH},
the angular momenta of the microscopic constituents, recently also observed \cite{EHsingle}, add up to a macroscopic angular momentum that will set the object in rotation.
We will find that the electromagnetic quantum fluctuations of the field modes of the Mie sphere participate in the conservation of angular momentum, and emerge as a  {QED} correction to
the diamagnetic  Einstein-De Haas effect  {of the Mie sphere as a whole}. This correction depends explicitly  {on the  microscopic model }for
the optical  {and magneto-optical }constants of the sphere at all frequencies.

 \section{Lifshitz theory of angular momentum}

 {Electromagnetic quantum fluctuations are first of all electromagnetic fields. Their electric fields at frequency $\omega$ satisfy the Vector Helmholtz equation \cite{jackson,cohen},}
 \begin{eqnarray}\label{VHH}
 \left[ - \bm{\nabla} \times \bm{\nabla} + \bm{\varepsilon}(\mathbf{r}, \omega)  \frac{\omega^2}{c_0^2} \right] \mathbf{E}(\mathbf{r},\omega) = 0
 \end{eqnarray}
{ The correlation function of }{electromagnetic quantum} fluctuations  obeyes the  Fluctuation-Dissipation Theorem, which basically assumes quantum excitations to be in equipartition throughout
 {available} phase space. The Lifshitz theory applies this  {principle} to a dielectric medium
in which case   { the spectral density of} the electric field is given by \cite{LLstat2}
\begin{eqnarray}\label{fd}
    \langle \bar{E}_i(\mathbf{r}_1,\omega) E_j(\mathbf{r}_2,\omega) \rangle d\omega  &=&   \frac{4 \omega Q(\omega) d\omega }{-2ic_0^2}
    \nonumber \\
&&  (\Delta \mathbf{G})_{ij}(\mathbf{r}_1,\mathbf{r}_2,\omega+i0)
 \end{eqnarray}
with the anti-hermitian operator $   \Delta \mathbf{G} \equiv \mathbf{G}-  \mathbf{G}^\dag$ in terms of
 the classical (retarded)  Green's tensor $\mathbf{G}$ associated with  the Vector Helmholtz wave equation~(\ref{VHH}) ({from now on called the
 ``Helmholtz Green tensor''} ) for the electric field. We use the symbols $\dag$ and  $\overline{\cdots}$ for full hermitian conjugate and complex conjugation, respectively.
  {The expectation value is taken for a quantum state of matter and photons in thermal equilibrium at temperature} $T$ .
  {At zero temperature such a state involves random quantum fluctuations in the occupation of
 electromagnetic modes of the  matter at all frequencies with the  energy $Q(\omega) = \hbar \omega$ for
 positive frequencies and $Q(\omega) = 0$ for negative frequencies.

 {The great advantage of the Lifshitz approach} \cite{milonni,lifshitz}  {is that it facilitates the description of microscopic electromagnetic quantum fluctuations in macroscopic matter, described by macroscopic Maxwell equations.
Van Enk} \cite{enk}  {applied Lifshitz
theory to calculate the electromagnetic angular momentum of the quantum vacuum between two uniaxial dielectric plates.}

For a genuine, infinite quantum vacuum at zero temperature, the electromagnetic eigenmodes are plane waves with wave number
$\mathbf{k}$ and
two degenerate transverse polarizations.  {Their average momentum and angular momentum both vanish, and}  Eq.~(\ref{fd})  {for their average energy density reduces to the familiar
diverging expression }
\begin{equation*}
   \sum_\mathbf{k,k'}  \int_0^\infty d\omega   \left< \frac{\bar{E}_i(\mathbf{k},\omega)E_i(\mathbf{k}',\omega )}{4\pi } \right> = 2 \sum_{\mathbf{k}}\frac{1}{2}\hbar \omega_\mathbf{k}
\end{equation*}
with $\omega_\mathbf{k} = kc_0$.  {For a macroscopic dielectric sphere  the eigenfunctions are described by vector spherical harmonics with angular momentum} $M \hbar$.

The
rigorously conserved quantity in   Maxwell-Lorentz  theory is
 $\mathbf{J}_{\mathrm{kin}} +  {\mathbf{J}}$ \cite{bartepj}, with $\mathbf{J} = \int d^3 \mathbf{r}\,  \mathbf{r} \times\mathbf{ K}$
 and momentum density  $ \mathbf{K} =
\mathbf{\bar{E}}\times \mathbf{B}/4\pi c_0$.
Under very general circumstances,
we can establish from Eq.~(\ref{fd}) that \cite{bartepj},
\begin{eqnarray}\label{noPoyn}
 \bm{\nabla} \cdot \left \langle \mathbf{\bar{E }}(\mathbf{r})\times\mathbf{ H}(\mathbf{r}) \right\rangle &=&0 \nonumber \\
     \int d^3 \mathbf{r} \, \left \langle \mathbf{\bar{E }}(\mathbf{r})\times\mathbf{ H}(\mathbf{r}) \right\rangle &=& 0
\end{eqnarray}
This  {first} expression states that  {electromagnetic fluctuations obeying }Eq.~(\ref{fd}) do not move energy.  The  {second }expression implies that
the total average electromagnetic momentum,
$\int d^3\mathbf{ r}\, \langle \mathbf{K}(\mathbf{r}) \rangle $ vanishes, at least when $\mu(\omega)=1$ so that  $\mathbf{H}=\mathbf{B}$. As a result,  the average electromagnetic
angular momentum is independent on the choice of the origin.
The Poynting vector itself does not need to vanish, for instance, the curl of some vector field satisfies both equations~(\ref{noPoyn}).

 {Following }Ref.~\cite{cohen} the quantum expectation
of the electromagnetic angular momentum  is split up into  a part with a transverse electric field ($\bm{\nabla} \cdot \mathbf{E}_T \equiv  0$) and one with a  longitudinal electric field ($\bm{\nabla} \times\mathbf{ E}_L \equiv 0$).
 {The transverse  electric field - in the Coulomb gauge determined by the time derivative of the vector potential $\mathbf{A}(\mathbf{r},t)$ - is  identified with photons with transverse polarization whose electromagnetic angular momentum is described by
the angular momentum operator} $\mathbf{L} + \mathbf{S}$  {representing orbital angular momentum and spin. In the same Coulomb gauge, the longitudinal electric field is
described by the spatial gradient of
a scalar potential } $\phi(\mathbf{r},t)$.  {As was seen in section I, the longitudinal electric field  generates an angular momentum }
related to the vector potential $\mathbf{A}$.
This ''longitudinal" angular momentum adds up to the kinetic angular momentum to get the canonical angular momentum, introduced earlier in Eq.~(\ref{QMlenz})
\cite{cohen}.
 {By working out the total electromagnetic angular momentum } $\int d^3\mathbf{ r}\, \mathbf{r}\times \langle \mathbf{K}(\mathbf{r}) \rangle_z $
 using Eq.~(\ref{fd}),  {the three contributions $\mathbf{S}$, $\mathbf{L}$ and $\mathbf{J}_{\mathrm{long}}$ arise after some algebra}
 \cite{cohen},

\begin{eqnarray}\label{jz}
   J_z  &=& -\frac{1}{2\pi i c_0^2} \int_{-\infty}^{\infty}   {d\omega \, Q(\omega)} \mathrm{TR} \, \mathbf{J}_z
  \cdot [ \mathbf{G}(\omega) - \mathbf{G}^\dag(\omega) ]  \nonumber \\
\end{eqnarray}
From now on we drop the brackets denoting quantum
average.  {The trace $\mathrm{TR}$ stands for the full  trace
in classical electromagnetic Hilbert space, to be specified below. The angular momentum operator acts in the same Hilbert space, and is given by}
$\mathbf{J}_z = \mathbf{L}_z + \mathbf{S}_z + \mathbf{J}_{\mathrm{long}, z} $ featuring orbital
momentum $(\mathbf{L}_z)_{mn} = (\mathbf{r} \times\mathbf{ p})_z
\delta_{mn}$,  spin $(\mathbf{S}_z)_{mn} = -i\epsilon_{zmn}$, and longitudinal angular
 momentum  $(\mathbf{J}_{\mathrm{long},z})_{mn} = -\epsilon_{zmk} r_k p_n$.
  {In classical Hilbert space,  it is convenient to introduce the operators }$\mathbf{p} = -i \bm{\nabla} $ and $\mathbf{S}_z$
   {without the usual factor} $\hbar$.

The Helmholtz Green{ tensor} is determined from Eq.~(\ref{VHH}) and given by the operator,
\begin{eqnarray}\label{hh}
    \mathbf{G}(\mathbf{r},\mathbf{p},z) &= &\frac{1}{\bm{\varepsilon}(\mathbf{r},z)z^2/c_0^2 - p^2\mathbf{ \Delta}_p }
\end{eqnarray}
with the transverse operator $p^2(\bm{\Delta}_p)_{nm} = p^2 \delta_{nm} - p_np_m$, and is - by causality- analytic for all complex frequencies $z$  with $\mathrm{Im}\, z > 0$.
For an  {external, slowly varying} magnetic field  {pointing} in the $z$ direction, the dielectric tensor {with magneto-optical activity linear in the magnetic field} is
\begin{equation}\label{FE}
    \varepsilon_{nm}(\mathbf{r},z) = m^2(r,z) \delta_{nm} -  W(r,z) \frac{c_0}{z}(\mathbf{S}_z)_{nm}
\end{equation}
in terms of the hermitian spin operator defined above.
 {The symmetries }$P$,$C$ and $T$  {allow the existence of an angular momentum of a homogeneous {magneto-optically active} sphere,
 linear in the {magnetic field. }}
{The non-perturbational solution of the Mie problem with magneto-anisotropy described by Eq.~(\ref{FE}) is known in literature \cite{NPMie}. The typical small parameter
in perturbation theory is $W(k)/k$ (with $k = 2 \pi/\lambda$ the wave number, roughly equal to the amount of Faraday rotation per wavelength. Strong Faraday rotation at
optical frequencies and magnetic
fields of $1$ Tesla is typically
equal to $200$ rad/m  and thus as small as  $10^{-4}$ over one wavelength of $0.5$ $\mu$m. This small number justifies us} to treat the magneto-optical effect
$W(r,z)$ as a linear perturbation to the standard Mie problem \cite{lacoste}, and Eq.~(\ref{hh}) is expanded as ,
 \begin{eqnarray}\label{hhper}
    \mathbf{G}(z) &= & \mathbf{G}_M(z) + \mathbf{G}_M(z) \cdot  \frac{z}{c_0}W(r,z) \mathbf{S}_z \cdot \mathbf{G}_M(z)
\end{eqnarray}
in terms of the Green's tensor  $\mathbf{G}_M$ of the  Mie sphere without magnetic field.  {It can be expressed in terms of the complete set of Mie eigenfunctions }\cite{newton}.
 {Expressions}~(\ref{jz}) and (\ref{hhper})  { relate the angular momentum of the sphere
induced by the magnetic field, and thus the Einstein-de Haas effect,
to the optical magneto-dichroism  and the full set of Mie eigenfunctions at all frequencies.}

A rigorous  microscopic theory for molecular
magneto-dichroism is given by Barron \cite{barron} and contains both paramagnetic and diamagnetic contributions.  {In this work we will adopt the
simplest description of the {magneto-optical }effect, one that is due to a uniform Zeeman splitting of all involved optical transitions and associated with the so-called diamagnetic ``A-term" in the rigorous  microscopic theory.}
In that case, $ \bm{\varepsilon}(\omega) = m^2(\omega + \omega_c\mathbf{ S}_z) $ so that  $W = -(\omega_c/c_0) \omega  dm^2/d\omega $,  {a relation that is attributed to Becquerel }
\cite{serber}.
 {In the next sections we verify first that} Eq.~(\ref{jz}) reproduces the angular momentum~(\ref{ho}) of a single harmonic quantum oscillator.
 {Next, we investigate how the collective angular momentum of the sphere is modified by the Mie eigenfunctions}.

\section{Single electric dipole}\label{dipolesection}
The classical, {phenomenological } model of an harmonically bound electron interacting with an electromagnetic wave of frequency $\omega$ and a (quasi-) static external magnetic field $\mathbf{B}_0$
{chosen along the $z$-axis}  is described by the equation
\begin{equation}\label{classdipole}
    \frac{ d^2\mathbf{r}}{dt^2} = -\omega_0^2 \mathbf{r }  + \frac{e}{m} \mathbf{E}(t) + \frac{e}{mc_0}  \frac{d\mathbf{r}}{dt}\times
    \mathbf{B}_0 - A  \frac{d\mathbf{r}}{dt}
\end{equation}
The last term describes the radiation loss with {time-scale} $ 1/A \sim c_0 / r_e \omega_0^2\tau_R $ \cite{jackson}.
For an harmonic field $\mathbf{E}(\omega ) \exp(-i\omega t) $
the electric dipole moment $\mathbf{d}= e \mathbf{r}$  satisfies $\mathbf{d}(\omega) =
\bm{ \alpha}(\omega) \cdot \mathbf{E}(\omega)$ with the polarizability given by
\begin{equation}\label{alpha1}
    \bm{\alpha}(\omega) = \frac{e^2/m}{\omega_0^2 -(\omega+\omega_c \mathbf{S}_z)^2 - iA \omega }
    \end{equation}
with $\omega_c$ the cyclotron frequency {introduced earlier and negative for the electron}.{ The most severe failure of this model is the divergence of the frequency moments higher than the second of its imaginary part, but this causes
no problem in the present work \cite{ChL}.}   {The small parameter that justifies a linear
expansion of the magneto-optical effect in this polarizability is $\omega_c/\omega_0$.  Since
$|\omega_c| = 15$ MHz for a magnetic field of one Tesla,
this parameter is very small if we consider the optical resonance to occur in the visible region. }

We can compare this outcome to scattering theory that contains a singularity for large wave numbers that we will identify. The {Vector }Helmholtz Green{ tensor} (\ref{hh}) is related to the $T$-matrix of the dielectric object according to,
\begin{eqnarray*}
    \mathbf{G}(\mathbf{k},\mathbf{k}',\omega ) =  \mathbf{G}_0(\mathbf{k},\omega ) \delta_{\mathbf{kk'}} +  \mathbf{G}_0(\mathbf{k},\omega )\cdot \mathbf{T}_{\mathbf{kk}'}(\omega)
     \cdot   \mathbf{G}_0(\mathbf{k}',\omega )
\end{eqnarray*}
with $ \mathbf{G}_0 (\mathbf{k},z) = \hat{\mathbf{k}}\hat{\mathbf{k}} c_0^2 /z^2 + \mathbf{\Delta}_k / [z^2/c_0^2 - k^2] $ the one of empty space, and $z = \omega+i0$ for the retarded Green's function.
Given an incident plane wave $\mathbf{E}_{\mathbf{kg}}(\omega)$ with wave number $\mathbf{k}$, frequency $\omega$, and transverse polarization $\hat{\mathbf{g}}$, the induced dipole moment of
the object is seen to be given by $\mathbf{d}
(\omega)= -\mathbf{T}_{\mathbf{0k}}(\omega)\cdot \mathbf{E}_{\mathbf{kg}}(\omega)  c_0^2/\omega^2$ \cite{pr}. Because the dipole is small compared to the wavelength,
we assume that the  $T$-matrix does  not depend on wave vectors, so that
$\mathbf{T}_{\mathbf{0k}}(\omega) := \mathbf{t}(\omega) =  -\bm{\alpha}(\omega) \omega^2/c_0^2$.
We can now evaluate the trace in Eq.~(\ref{jz})  { using plane waves }(where $\sum_\mathbf{k} \equiv \int d^3 \mathbf{k}/(2\pi)^3$  {having the
 physical unit $1/m^3$} )
\begin{eqnarray*}
     dL_z(\omega)   &+ &   dS_z(\omega)  =  \frac{ \hbar \omega d\omega}{2\pi c_0^2}  \epsilon_{znm} \\
    && \times  \sum_\mathbf{k}  \frac{1}{[\omega^2/c_0^2 - k^2]^2}
    \left[ \mathbf{ \Delta}_k \cdot \mathbf{t}(\omega) \right]_{nm} -\mathrm{h.c. }
\end{eqnarray*}
and
\begin{eqnarray*}
      dJ_{\mathrm{long},z}(\omega)   &&=  -\frac{ \hbar \omega d\omega }{2\pi c_0^2}\epsilon_{znm}  \\
    && \times  \sum_\mathbf{k} \frac{1}{\omega^2/c_0^2} \frac{1}{\omega^2/c_0^2 - k^2}
     \left[ \mathbf{\mathbf{t}(\omega) \cdot  \Delta}_k \right]_{nm} - \mathrm{h.c}.
\end{eqnarray*}
The first integral over $\mathbf{k}$ for the transverse angular momentum converges for all frequencies, whereas the  longitudinal angular momentum  suffers from a divergence of
 the type $\sum_\mathbf{k} 1/k^2$  {reminiscent of the Coulomb singularity}. In Appendix~\ref{appA} we show that the relation $\mathbf{t}(\omega) =  -\bm{\alpha}(\omega) \omega^2/c_0^2$  is consistent with a  regularization of
  this integral as  $3/{2r_e}$, with $r_e$ the classical electron radius.
 {  Physically this means that the} $k$- {dependence neglected earlier in the $T-$matrix, comes in at wave numbers as large as} $k \sim 1/r_e$,  {at least for
the transverse plane waves}.
The longitudinal angular momentum exhibits exactly the same transverse divergence and we shall deal with it in exactly the same way. The diverging part then becomes,
\begin{eqnarray}\label{Jcandiv}
    J_{\mathrm{long},z} &=& -\frac{ \epsilon_{znm} }{2\pi c_0^2 r_e} \int_{0}^\infty d\omega\, \hbar \omega \left(\alpha_{nm}(\omega) - \alpha^\dag_{nm}(\omega) \right) \nonumber \\
    &=& \hbar \times \frac{\omega_c}{\omega_0}
\end{eqnarray}
to leading order in $\omega_c/\omega_0$.
 The remainder, including the transverse angular momentum, has a finite $k$-integral proportional to $ic_0/\omega$ and we find
\begin{equation}\label{Jremain}
    J_{\mathrm{rem},z} \sim   \frac{{\hbar \omega_c r_e}}{c_0} \mathrm{ Re}\, \int_{0}^\infty d\omega\,  \frac{\omega^3}{\left[\omega_0^2 -(\omega + iA/2)^2\right] ^2 }
\end{equation}
In this case the frequency integral diverges logarithmically, much like the non-relativistic derivation of the Lamb-shift \cite{milonni}.
We will attribute this modest singularity
at  high-frequencies to the failure of the present
model, and assume  that it is eliminated by  QED \cite{kawka} {beyond frequencies of order $ mc_0^2/\hbar$.}
Replacing the frequency integral by a  number of {order $\log  mc_0^2/\hbar \omega_0  \sim 12 $ for optical transitions}, we infer that $J_{\mathrm{rem},z} $
differs from
$J_{\mathrm{long},z} $  by a factor $12 \times  \omega_0 r_e/c_0 \sim 10^{-6}$.

We conclude that the angular momentum induced
by the interaction of the quantum  {fluctuations} with the classical electric dipole
is entirely  associated with the longitudinal electromagnetic field. Equation~(\ref{Jcandiv})  {provides a relation between this angular momentum and the magneto-optical activity at all
frequencies. For the harmonic oscillator this} is equal to $\hbar \omega_c/\omega_0 $, which coincides with the outcome~(\ref{ho}) for a quantum mechanical oscillator,
who has this angular momentum is  ``built in".
In Ref.~\cite{milonni} this somewhat surprising coincidence is referred to as
''the necessity  for the vacuum field", and is linked to the fluctuation-dissipation relation. Similarly, for the single dipole an  energy
$E = \frac{3}{2}\hbar \omega_0$  {of the electromagnetic quantum fluctuations }can be
calculated from the fluctuation-dissipation theorem, which is precisely the energy of the quantum-mechanical ground state.  {According to Ref.~\cite{milonni}  this coincidence is meant to be,
and is here seen to be consistent with a regularization scheme that always regularizes
the same divergence in the same way.}

\section{Dielectric Mie Sphere}\label{miesection}
In this section we will express the angular momentum of a dielectric sphere with homogeneous  but frequency-dependent index of refraction inside, in terms of the Mie eigenfunctions.
A standard procedure in treatments of Casimir energy in macroscopic  media is to transform the frequency integral  {to a discrete sum of } Matsubara frequencies \cite{milton}, facilitated by the analycity
of $\mathbf{G}(z) $.
Since
 $W(z)$, like $\mathbf{G}(z)$, typically decays as  $1/z^2$ at large frequencies,  the frequency integral of
$\mathbf{G}(\omega+i0)$ can
be moved to the positive imaginary axis, and the one for $\mathbf{G}^\dag (\omega +i0) = \mathbf{G} (\omega - i0)$ to the negative imaginary axis.
One has to be careful to miss
neither longitudinal
poles in this expression at $\omega =0$ nor possible diverging contributions as $\omega \rightarrow \infty$. With this in mind \cite{notegz}, Eqs.~(\ref{jz}) and (\ref{hhper})  {transform into},
  \begin{eqnarray}\label{matsu}
  J_z = \frac{kT}{c_0^3 }\sum_{n=-\infty}^{\infty} &\, & s_n^2 W(is_n) \mathrm{TR} \,  \nonumber \\
&& \mathbf{J}_z \cdot \mathbf{G}_M(is_n)\cdot  \theta_a \mathbf{S}_z
  \cdot \mathbf{G}_M(is_n) \end{eqnarray}
with the Matsubara frequencies $s_n = 2 \pi n kT/\hbar$ and $\theta_a$  {putting the boundary at} $r=a$. The advantage of this formalism is that $\mathbf{G}_M(is_n)$
has become an hermitian operator, and  that both  $W(is_n)$ and $m(is_n)$ have become real-valued. The {Vector} Helmholtz operator for the
Mie problem without magnetic field is

\begin{equation}\label{hhmie}
    \mathbf{H}_{\mathrm{M}}(s)= {\varepsilon(r,is)}^{-1/2}p^2 \mathbf{\Delta}_p \, {\varepsilon(r,is)}^{-1/2}
\end{equation}
Since $s$ is real-valued, the operator $\mathbf{H}_\mathrm{M}(s)$ is
hermitian for all $s$. For any given $s$ it has orthonormal
eigenvectors written as $\left|\bm{\Psi}_{\nu JMk}(s) \right\rangle $ ($J$ and $M$ are the usual eigenvalues of the operators
$\mathbf{J}^2$ and $J_z$,   $\nu= e,m,\ell$
is a polarization index with prescribed parity, and $k$ is a wave number associated with radial decay).
They have
eigenvalues $K_{\nu k}^2=k^2 \geq 0$ for the two transverse polarizations $\nu=e,m$ and a highly degenerated eigenvalue $K_{\ell k}=0$
for the longitudinal eigenfunctions $\nu=\ell$ and all $k > 0$. For any given real-valued $s$, the
 wave functions
$\{\left|\bm{\Psi}_{\nu JMk}(s)\right\rangle\} $, including the longitudinal states,
constitute a complete orthonormal set that spans the electromagnetic Hilbert space,
equipped with a scalar product associated with
electromagnetic energy.
The  Helmholtz Green's {tensor} $\mathbf{G}_M$ of the Mie sphere  {featuring }in Eq.~(\ref{hh}) can now be decomposed as
\begin{eqnarray*}
&&  \mathbf{G}_M(is) =   {\varepsilon}^{-1/2}(is) \sum_{\nu JMk}   \frac{\left|\bm{\Psi}_{\nu JMk}\right\rangle \left\langle \bm{\Psi}_{\nu JMk}\right|   }{-s^2/c_0^2 -K_\nu^2}
 \,  {\varepsilon}^{-1/2}(is) \nonumber \\
 &=&  \sum_{JMk}   \frac{\left|\bm{E}_{\ell JMk}\right\rangle \left\langle \bm{E}_{\ell  JMk}\right|   }{-s^2/c_0^2} +
 \sum_{\nu=e,m\,  JMk}   \frac{\left|\bm{E}_{\nu JMk}\right\rangle \left\langle \bm{E}_{\nu JMk}\right|   }{-s^2/c_0^2 -k^2}
\end{eqnarray*}
where $\sum_k \equiv \int_0^\infty dk$ { (without here the usual factor $1/2 \pi$ for notation convenience)} . The electric field modes
$ \left|\bm{E}\right\rangle=  \varepsilon (r,is)^{-1/2}\left|\bm{\Psi}_{\nu JMk}\right\rangle$   appear that are  not orthonormal for different $k$.
For the transverse modes  $\nu = e,m$ with finite spin the sum over $J$ starts at $1$. For a Mie sphere with size $a$ and index of
refraction $m$ we write $\varepsilon (r,is) = 1 + (m^2(is) -1) \theta (a-r)$.
For an ideal Mie sphere,
$\theta(a-r)$ is the perfect step function on its surface, but in view of the many discontinuities that accumulate near the surface we shall assume it to be "smooth but rapidly
varying". We choose the  Lorenz-Lorentz model,  {found in many text books} \cite{feynman,vdhulst,jackson}

\begin{equation}\label{miediel}
    m^2 (is) = 1 +\frac{ \rho  \alpha (is)}{1- \frac{1}{3}  \rho  \alpha (is) }
\end{equation}
with $\rho \alpha(is) = \omega_p^2/(\omega_0^2 + s^2)$.  This models the sphere as a collection of harmonic oscillators with uniform volume density $\rho$, plasma
frequency $\omega_p^2 =\rho \alpha(0) \omega_0^2$, and with local field correction.

{Despite its simplicity the Lorenz-Lorentz model has no fundamental  shortcomings. It is analytic in the upper complex frequency plane provided that $\omega_p < \sqrt{3}\omega_0$, with correct limits
at low and high frequencies. Nevertheless, this excludes the consideration of too large plasma frequencies. Its only serious drawback is the
diverging high moments of its imaginary part mentioned earlier \cite{ChL}.
After Wick transformation to the imaginary frequency axis, the resonant poles are far from the integration axis, and the role of the dissipative part becomes negligible.}
{In this model, the static
dielectric constant is given by $m^2(0) = (1 + \frac{2}{3} \omega_p^2/\omega_0^2)/ (1 - \frac{1}{3} \omega_p^2/\omega_0^2) $ and one must have  again $ \omega_p < \sqrt{3} \omega_0 $
to have it finite at all frequencies. Close to this critical value local field corrections become large. In this work we consider  $ \omega_p/\omega_0 < 1.5$ and we realize that the
Lorenz-Lorentz model may cease to apply for even lower values of  the plasma frequency. }
 {The Lorenz-Lorentz model also excludes atomic correlations as well as recurrent electromagnetic scattering between dipoles that becomes relevant near atomic resonances\cite{pr}.
Both determine the Casimir energy of electromagnetic quantum fluctuations, but unlike for Casimir energy, divergencies in the angular momentum do not
originate from singular dipole-dipole coupling at small distances.
Also the recent Casimir puzzle about a possible inconsistency of the Drude model at low frequencies with observations of the Casimir force} \cite{Klimchtiskaya}
 {does not affect the present approach, since metallic behavior ($\omega_0$ close to zero) } is ruled out by our model, since we assume from the start the electrons
 to be bound.

It is straightforward to add more microscopic resonances to this model
but this is beyond the scope of this work. If the atomic resonance $\omega_0$ is located  in the visible regime, the effects of finite temperature
due to the discrete Matsubara sum happen beyond several thousand of degrees and will not be discussed.

\subsection{Transverse angular momentum}
The sum of orbital and spin momentum is governed by transverse modes so that in the following we consider
$\nu= e,m$ only \cite{cohen}.
Because $\left|\bm{\Psi}_{\nu JMk_\nu}\right\rangle$ is an eigenfunction of
the angular momentum operator  $\mathbf{L}_z+\mathbf{S}_z$ with eigenvalue $M$,  {we obtain from} Eq.~(\ref{matsu}),
\begin{eqnarray*}
&&    L_z  +  S_z  = \frac{2kT}{c_0^3}\sum_{n=1}^\infty W(is_n) s_n^2 \sum_{JM\nu}\sum_{kk'} M \\
\ \ \ &&  \ \  \times  \frac{ \left\langle \bm{\Psi}_{\nu JMk'}\right|  \varepsilon(r)^{-1}   \left|\bm{\Psi}_{\nu JMk}\right\rangle  \left\langle \bm{\Psi}_{\nu JMk'}\right|
 \varepsilon^{-1} \theta_a \mathbf{S}_z  \left|\bm{\Psi}_{\nu JMk}\right\rangle  }{(s_n^2/c_0^2 +k^2)(s_n^2/c_0^2 +k'^2)}
\end{eqnarray*}
Using $ \varepsilon(r)^{-1} = 1 +  (\varepsilon(r)^{-1} -1)$ we split this expression up into one term that counts angular momentum everywhere in space, and a second term that exists only
inside the sphere. The first simplifies using the orthonormality  of the eigenfunctions  $\left| \bm{\Psi}_{\nu JMk}\right\rangle$,
\begin{eqnarray*}
  (L_z  +  S_z)^{(1)} & = & \frac{2kT}{c_0^3}\sum_{n=1}^\infty W(is_n) s_n^2 \sum_{\nu JM k} M \\
\ \ \ &&  \ \  \times  \frac{  \left\langle \bm{E}_{\nu JMk}\right|  \theta_a \mathbf{S}_z  \left|\bm{E}_{\nu JMk}\right\rangle  }{(s_n^2/c_0^2 +k^2)^2}
\end{eqnarray*}
The matrix element involving the spin operator $\mathbf{S}_z$ is equal to $ \mathcal{E}_{\nu Jk}(a)  \times M/J(J+1)$ with
\begin{equation}\label{energiemie}
    \mathcal{E}_{\nu Jk}(a) \equiv \int_{r<a} d^3\mathbf{r} \,\left|\mathbf{E}_{\nu Jk}(\mathbf{r})\right|^2
\end{equation}
proportional to the electric energy of the mode $\{\nu JMk\}$  {at Matsubara frequency} $is$ inside the Mie sphere, and independent of $M$ \cite{lacoste}.
It can be large near specific resonant values for $k$
but this is of no relevance here since the integral over $k$  is not sensitive to narrow, large spectral peaks. With $\sum_{M} M^2= (2J+1) J (J+1)/3$ this gives the expression,
\begin{eqnarray*}
   (L_z  +  S_z)^{(1)}   = \frac{2kT}{3c_0^3}\sum_{n=1}^\infty W(is_n) s_n^2 \sum_{kJ\nu} \frac{ (2J+1) \mathcal{E}_{J\nu k}(a)}{(s_n^2/c_0^2 +k^2)^2}
\end{eqnarray*}
This Matsubara sum   {will be shown to} diverge logarithmically. This divergence is  more modest than the
quadratic divergence of Casimir energy \cite{milonni} of a dielectric sphere
or the Casimir momentum density in magneto-electric  media \cite{feigel}, but still it is diverging.
 {For large $s_n$ the refractive index  of the sphere is close to $1$ and the Born approximation applies}. The  wave functions are approximately equal to the transverse modes of free space,
\begin{eqnarray}\label{magneticmode}
    \bm{E}^{0}_{mkJM}(\mathbf{r}) &=&  \frac{4\pi i^J }{(2\pi)^{3/2}}k j_J(kr) \mathbf{\hat{X}}_{JM}(\mathbf{\hat{r}})\nonumber\\
      \bm{E}^{0}_{ekJM}(\mathbf{r}) &=& \frac{4\pi i^{J-1} }{(2\pi)^{3/2}} \frac{1}{r}\left[ \sqrt{J(J+1)} j_J(kr) \hat{\mathbf{N}}_{JM}(\hat{\mathbf{r}}) \right. \nonumber \\
      && \ \ \  +
      \left. \left(kr j_J(kr) \right)' \hat{\mathbf{Z}}_{JM}(\hat{\mathbf{r}}) \right]
\end{eqnarray}
in terms of the three orthonormal vector harmonics $\hat{\mathbf{X}}_{JM}$, $\hat{\mathbf{N}}_{JM}$ and $\hat{\mathbf{Z}}_{JM}$ \cite{cohen}.
 Their
 electric energy will be denoted by
$\mathcal{E}^{(0)}_{\nu Jk}(a)$.
When this energy is used rather than the full Mire solution, the angular momentum $(L_z+S_z)^{(1)}$ is determined {by electromagnetic quantum fluctuations} that
interact only with the {magneto-optical interaction} $W$
in the sphere without any more
scattering from the Mie sphere.
{Using the two summations~(\ref{sum1}) and (\ref{sum2})  derived in Appendix~\ref{appB}}
we can see that the remaining $k$-integral in the expression for $(L_z  +  S_z)^{(1)} $ equals $Vc_0/2s $ for both  polarizations, with $V$ the volume of the
Mie sphere. With $W \sim 1/s^2$ for large $s_n$, the  {Matsubara sum for} $(L_z  +  S_z)^{(1)} $ thus diverges logarithmically.  The remaining angular momentum,
associated with the stored energy
$\mathcal{E}_{\nu Jk}(a) - \mathcal{E}^{(0)}_{\nu Jk}(a)$ of scattered electromagnetic quantum fluctuations, is finite and depends smoothly on
the two parameters $\omega_p/\omega_0$ and $\omega_0a/c_0$. The same is true for the {third} term proportional to $  (\varepsilon^{-1}(is) -1)$ that was split off earlier,
and that vanishes fast enough  at large $s$ to make the Matsubara sum over $s_n$ converge. This is summarized at $T=0$ by

\begin{eqnarray}\label{LplusS}
  L_z  +  S_z  \approx  \frac{2\hbar N \omega_c r_e  }{3\pi c_0}  \left[
 \int_{\omega_0}^\infty \frac{ds}{s}   +  f_{LS}\left(\frac{\omega_p}{\omega_0}, \frac{\omega_0 a}{c_0}\right) \right]
\end{eqnarray}
with $f_{LS}$ a smooth function of order unity and $N = 4\pi a^3 \rho/3 $ the number of dipoles in the sphere.
The logarithmic divergence at high frequencies stems from the one  already encountered in Eq.~(\ref{Jremain}) for a
 single oscillator, and was already argued to be irrelevant.
Therefore, the
electromagnetic angular momentum due to orbit and spin
induced by the quantum  {fluctuations i}s of order $ N \times \hbar\times \omega_c r_e/c_0 $.
The last factor $\approx 10^{-16} $ makes the transverse angular momentum entirely negligible .

\subsection{Longitudinal angular momentum}

The longitudinal angular momentum is  determined by the longitudinal electric field of the { electromagnetic quantum fluctuations}.
In the absence of{ magneto-optical activity}, the
longitudinal  {electric modes}  exist only at the surface $r=a$ of the sphere \cite{newton, vdhulst}. The spin operator $\mathbf{S}_z$
associated with the magneto-optical perturbation~(\ref{hhper})
couples transverse electric modes ($\nu = e$) to longitudinal
modes inside the sphere.

Any  longitudinal Mie eigenfunction of $ \mathbf{H}_{\mathrm{M}}$ defined in Eq.~(\ref{hhmie}) can be written as
$ \left|\bm{\Psi}_{\nu=\ell, JMk}\right\rangle  =
 - k^{-1}\varepsilon(r)^{1/2} \mathbf{p} \left| \Phi_{kJM} \right\rangle $, with $\Phi_{kJM} (\mathbf{r})$ essentially the scalar potential field,
 the factor $k$ introduced for
 reasons of normalization and dimension.
 Since $[\mathbf{L}_z +\mathbf{S}_z, p_i ] =0$ this is an
 eigenfunction of $ \mathbf{L}_z +\mathbf{S}_z$ provided that
$ \left| \Phi_{kJM} \right\rangle $ is an eigenfunction of $L_z$.   To have normalized longitudinal eigenfunctions, we require that
\begin{equation*}
  \left\langle \bm{ \Psi}_{\ell k'J'M'}\right.  \left| \bm{\Psi}_{\ell kJM} \right\rangle  =  \delta(k-k')\delta_{MM'} \delta_{JJ'}
 \end{equation*}
This is satisfied if the scalar potentials are the orthonormal solutions of the  eigenvalue equation,
\begin{equation}\label{long}
    \mathbf{p} \cdot \varepsilon(r) \mathbf{p} \left| \Phi_{kJM} \right\rangle  = k^2 \left| \Phi_{kJM} \right\rangle
\end{equation}
normalized such  that they behave far outside the sphere as $ (4\pi/(2\pi)^{3/2})\sin (kr-\phi_J) /r $ \cite{merz}.
The longitudinal angular momentum becomes ($ \kappa := \{kJM\}$),
\begin{eqnarray}
  J_{\mathrm{long},z} & =& -\frac{2kT}{c_0}\sum_{n=1}^\infty W(is_n)\sum_{\kappa}\sum_{\kappa'}\sum_{\nu =e/m}   \nonumber \\
&&   \frac{ \left\langle \bm{E}_{\nu \kappa'}\right| \cdot ( \bm{\epsilon}_z\cdot \mathbf{r}) \mathbf{p}^2    \left| \Phi_{\kappa} \right\rangle
 \left\langle \Phi_{\kappa} \right| \mathbf{p}\cdot \theta_a \mathbf{S}_z  \cdot\left|\bm{E}_{\nu \kappa'}\right\rangle  }{k^2 (s_n^2/c_0^2 +k'^2)} \label{jlongz}
\end{eqnarray}
We inserted $\left|\bm{E}_\kappa \right\rangle =
\varepsilon (r,is)^{-1/2} \left|\bm{\Psi}_\kappa\right\rangle $ for the transverse eigenfunctions.
The sum over $\nu$ runs only over the two transverse modes since longitudinal modes alone cannot generate angular momentum.
We will first ignore a surface contribution, use Eq.~(\ref{long}) away from  {the dielectric discontinuity at} $r=a$  and use  $\mathbf{p}^2 \left| \Phi_{\kappa} \right\rangle = k^2\varepsilon^{-1}
 \left| \Phi_{\kappa} \right\rangle$. Using the closure relation $\Sigma_\kappa \left| \Phi_{\kappa} \right \rangle \left\langle \Phi_{\kappa} \right|=1$,
\begin{eqnarray*}
  J^{(1)}_{\mathrm{long},z}  &=& -\frac{2kT}{c_0}\sum_{n=1}^\infty W(is_n)     \times  \\
&&   \sum_{\kappa ,\nu =e/m} \frac{ \left\langle \bm{E}_{\nu \kappa }\right| \cdot ( \bm{\epsilon}_z\cdot \mathbf{r}) \varepsilon^{-1} \mathbf{p}\cdot  \theta_a \mathbf{S}_z \cdot
\left|\bm{E}_{\nu \kappa}\right\rangle  }{s_n^2/c_0^2 +k^2}
\end{eqnarray*}

This expression can be further simplified by averaging over de orientation of the $z$-axis and by
using $ \mathbf{p} \theta_a(r)  = -i\hat{\mathbf{r}} \partial_r \theta_a  +   \theta_a \mathbf{p} $.
{The surface terms created by the radial derivative $\partial_r \theta_a$ complicate the analysis.} To treat discontinuities across the surface in the term involving $\partial_r \theta_a$, we imagine the surface profile $\varepsilon(r) = 1 + (m^2-1) \theta_a$ to be a smooth function that decays from $1$ to $0$
 over a small range $2w$ around $r=a$ and use
\begin{equation}\label{truc}
    \int_{a-w}^{a+w} dr f( \theta_s)\frac{ d\theta_s}{dr} = F(\theta_a = 0) - F(\theta_a = 1)
\end{equation}
with $F$ the primitive function of $f$. Because the electric field $\mathbf{r} \times \mathbf{E}$ directed along the surface of all modes is continuous for a
perfectly sharp surface,  it is approximately constant in the thin layer around  $r=a$. Thus,
$1/\varepsilon(\theta_a)= 1/(1 + (m^2 -1)\theta_a)$ is the only  factor in $ J^{(1)}_{\mathrm{long},z}$ that   varies significantly across the surface.
With Eq.~(\ref{truc}) giving $-(\log  m^2)/(m^2 -1)$, the surface contribution becomes
\begin{eqnarray*}
    J_{\mathrm{long},z}^{(1a)}  &=& \frac{2kT}{c_0} \frac{a^3}{3}\sum_{n=1}^\infty W(is_n)   \sum_{\kappa,\nu}  \frac{1}{s_n^2/c_0^2 + k^2}\times  \\
   && \ \ \frac{\log m^2}{m^2 -1  } \int d^2\mathbf{\hat{r}} \left|\mathbf{ E}^T_{\nu \kappa,z}(a\mathbf{\hat{r}})\right|^2
\end{eqnarray*}
 with $\mathbf{E}^T=\hat{\mathbf{r}} \times \mathbf{E} $ the electric field parallel to the surface.
 The term involving $\theta_a  (S_z)_{nm} p_n E_m$  {in the above expression for} $ J^{(1)}_{\mathrm{long},z} $ can be treated similarly, with the normal displacement vector $\varepsilon(r) E^L=
 \varepsilon(r) \hat{\mathbf{r}} \cdot \mathbf{ E} $ continuous across
 the surface. Adding it to $J_{\mathrm{long},z}^{(1a)} $ gives,
 \begin{eqnarray}\label{Jcan1}
  &&  J_{\mathrm{long},z}^{(1)}   = \frac{2kT}{c_0}\frac{a^3}{3}\sum_{n=1}^\infty W(is_n)   \sum_{\kappa,\nu}  \frac{1}{s_n^2/c_0^2 + k^2}\times  \nonumber  \\
   && \int d\mathbf{\hat{r}} \left\{T(m)  \left|\mathbf{ E}^T_{\nu \kappa}(a\mathbf{\hat{r}})\right|^2 +
   L(m)  { E}^L_{\nu \kappa}(a\mathbf{\hat{r}})^2 \right. \nonumber  \\
   && \ \ \  \left. + \frac{1}{2 m^2 a^3} \int_0^a dr\, r^2   \, \left|\mathbf{ E}_{\nu \kappa }(\mathbf{r})\right|^2 \right\}
\end{eqnarray}
where all the fields are defined  inside the sphere, and
\begin{eqnarray*}
  T(m) &=& \frac{\log m^2}{m^2-1} - \frac{1}{2m^2} \\
  L(m) &=& \frac{m^4 + m^2 +1}{3m^2} - (m^2-1) \\ && \ \ \ -  \frac{1}{2m^2} - \frac{(m^2-1)(m^2+2)}{6m^2}
\end{eqnarray*}
 determine the weight  of the transverse  and  normal components of the eigenfunctions at  the surface. Note that the first and third term in Eq.~(\ref{Jcan1}) are strictly positive, but
 that the normal components $E^L(a)$ at the surface can make a negative contribution when $m$ is large enough.

Finally, we have to  consider the
surface term ignored earlier,  {due to the discontinuity of} $\varepsilon(r)$  {at the surface of the sphere}, and identified as
\begin{equation*}
    \mathbf{p}^2 - \frac{1}{\varepsilon(r)} \mathbf{p}\cdot \varepsilon(r)\mathbf{p} = \frac{m^2-1 }{1 + (m^2-1)\theta_a } \left(i \partial_r \theta_a \right)  \frac{1}{i}\partial _r
\end{equation*}
Since the transverse electric field and
the normal, longitudinal  displacement vector
$-\varepsilon(r) \partial_r \Phi_{\kappa}$ are both constant over the thin layer, we identify the rapidly varying function
$f = - (m^2-1)/ \varepsilon(r)^2$  in the first matrix element of
the expression~(\ref{jlongz}) for $ J_{\mathrm{long},z}$.   Equation~(\ref{truc})
produces $1 - 1/m^2$ for the { radial integral across the surface}, and one obtains,
 \begin{eqnarray} \label{J2}
  &&  J_{\mathrm{long},z}^{(2)} = - \frac{2kT a^3}{c_0}\sum_{n=1}^\infty W(is_n)   \sum_{\kappa\kappa'}  \frac{(m^2 -1)}{s_n^2/c_0^2 + k^2}
   \times \nonumber  \\
   && \int_{r=a} d\mathbf{\hat{r}}   (i\bar{\mathbf{E}}^T_{e \kappa})_z {E}^L_{\ell\kappa'}
  \times \int_{r<a} d^3\mathbf{ r} \, (\bar{\mathbf{E}}_{\ell\kappa'}\cdot \mathbf{ S}_z \cdot  \mathbf{E}_{e\kappa} )
\end{eqnarray}
with all fields again to be evaluated inside the sphere, and where $\bm{\nabla} \Phi_\kappa = ik \mathbf{E}_{\ell\kappa}$ has been substituted.
Because of  parity, only the transverse electric mode $\nu = e$ couples to the longitudinal mode,  {and the mode} $\nu = m$  {is absent.}

We infer from Eqs.~(\ref{Jcan1}) and (\ref{J2})  that a significant amount of angular momentum can be expressed as an integral over the surface of the sphere.
This is to  be expected because
that is where  longitudinal electric fields and surface charges of the Mie sphere reside. However, the separation of angular momentum into bulk and surface is
mathematically not unique, only total angular momentum
is \cite{notU}.  Nevertheless,  Eq.~(\ref{Jcan1})
clearly contains a bulk term
as well, induced by longitudinal electric waves created by{ magneto-optical activity}. The
angular momentum $J_{\mathrm{long},z}^{(2)} $ resides at the surface as well but
originates from  a mode conversion of transverse electric to longitudinal electric field modes by the magneto-optical activity  inside the sphere, as can be induced  from the
second, bulk integral of Eq.~(\ref{J2}).
\bigskip

 {We will now evaluate Eqs.~(\ref{Jcan1}) and (\ref{J2}) and identify their divergencies. }Since $W(is) \sim 1/s^2$ for large $s$, no divergencies  show up in the frequency integral of the longitudinal angular momentum  {and the Matsubara frequencies }$s  \sim \omega_0$  {give the dominant contribution.}
However, $J_{\mathrm{long},z}^{(1)}$  and $J_{\mathrm{long},z}^{(2)}$  share the divergence of the  $k$-integral for all frequencies
and for all $J$. This UV catastrophe is not due to the broadband quantum fluctuations, and
 would even exist for a Mie sphere exposed to
a bandwidth-limited classical noise. It is caused by the breakdown of  the description of
the Mie sphere at the smallest length scale associated with Coulomb charges, which is the electron radius.
We \emph{postulate} that this divergence is regularized  {in exactly the same way} as was done in section~\ref{dipolesection}  for a single oscillator, { and justify this choice by discussing the end result}.

To extract the divergence we focuss on the Mie problem for fixed $s \sim \omega_0$ and large $k$. The transverse  Mie eigenfunctions inside the sphere are given by
\begin{eqnarray}\label{mieeigen}
\bm{E}_{mkJM}(\mathbf{r}) &=&  \frac{4\pi i^J }{(2\pi)^{3/2}}A^m_{Jk} k j_J(mkr) \mathbf{\hat{X}}_{JM}(\mathbf{\hat{r}})\nonumber\\
\bm{E}_{ekJM}(\mathbf{r}) &=& \frac{4\pi i^{J-1} }{(2\pi)^{3/2}}A^e_{Jk}\frac{1}{r}\left[ \sqrt{J(J+1)} j_J(mkr) \hat{\mathbf{N}}_{JM}(\hat{\mathbf{r}}) \right. \nonumber \\
&& \ \ \  + \left. \left(mkr j_J(mkr) \right)' \hat{\mathbf{Z}}_{JM}(\hat{\mathbf{r}}) \right]
\end{eqnarray}
They involve complex amplitudes $A^\nu_{Jk}(is)$ that depend on $m(is)$, $J $ and the size parameter $ka$ of the sphere.
{Their absolute values are imposed  by continuum normalization, similar to what was done earlier in Eq.~(\ref{long}) with the longitudinal modes.
 Also for the transverse eigenfunctions, normalization is equivalent to imposing the amplitude of the field mode far outside the Mie sphere.}

For $ka \rightarrow \infty$ we easily obtain from the
exact expressions
\cite{newton,vdhulst}
\begin{eqnarray*}
    \left|A^{m/e}_{Jk}\right|^2 & \rightarrow & \frac{2}{m^2 +1 + (m^2-1) \cos(2kma -J\pi )}
\end{eqnarray*}
If we ignore the oscillating function  {in this expression when performing }the integral over $k$, these asymptotic expressions no longer depend on $J$ and $k$.
If we also restrict to zero temperature, the contribution of the transverse magnetic mode ($\nu = m$)
to Eq.~(\ref{Jcan1}) is

\begin{eqnarray*}
 &&J_{\mathrm{long},z}^{(1,m)} = \frac{\hbar a^3}{3\pi  c_0 } \frac{(4\pi)^2 }{(2\pi)^3}   \int_0^\infty ds\, W(is)  \frac{2}{m^2+1 }\sum_{kJ} (2J+1) \\
    && \times \left( T(m) j_J^2(mka)  +\frac{1 }{2m^2 a^3}\int_0^{a} dr r^2 j_J^2(mkr)\right)
\end{eqnarray*}
We insert $ \sum_{J=1} (2J+1) j_J^2(y)= 1 - j_0^2(y)$, {equivalent to Eq.~(\ref{sum1}) of Appendix~\ref{appB}}, and ignore the oscillating term $j_0^2$ that
produces a  finite and negligible longitudinal angular momentum proportional to the surface $a^2$ of
the sphere.
What remains is,
\begin{eqnarray*}
  J_{\mathrm{long},z}^{(1,m)}  = \frac{ 4\hbar  a^3}{3 c_0 }  \frac{3 }{2 r_e} \int_0^\infty ds\,
  \frac{2W(is)}{m^2+1 } \left( T(m) + \frac{1}{6m^2}  \right)
\end{eqnarray*}
 where we have regularized, as in section~\ref{dipolesection}, $4\pi /(2\pi)^3 \int dk = 3 /2r_e$.
 The transverse electric wave can be treated similarly, this time using the sums (\ref{sum2}) and (\ref{sum3}). The result is ,
 \begin{eqnarray*}
  J_{\mathrm{long},z}^{(1,e)}  &= &\frac{ 4\hbar  a^3}{3 c_0 }  \frac{3 }{2 r_e} \int_0^\infty ds\,
  \frac{2W(is)m^2}{m^2+1 }  \\ &\times & \left( \frac{T(m)+ 2 L(m) }{3}+ \frac{1}{6m^2}  \right)
\end{eqnarray*}
Only for $m=1$ is $J_{\mathrm{long},z}^{(1,e)} = J_{\mathrm{long},z}^{(1,m)}$.  {For diamagnetism in the} simple model (\ref{miediel}) we find
\begin{equation*}
 W(is)= -2 \frac{\omega_c s^2}{c_0}  \frac{dm^2}{ds^2} = \omega_c \left(\frac{m^2+2}{3}\right)^2 \frac{ 2\rho r_ec_0 s^2}{(\omega_0^2 + s^2)^2}
\end{equation*}
 Upon substituting $x= s/\omega_0$, the total angular momentum for both polarizations $e$ and $m$ is,
 \begin{eqnarray}\label{jcanfinal}
  J_{\mathrm{long},z}^{(1)}  = N \hbar \frac{ \omega_c}{\omega_0}  && \frac{1}{\pi }\int_0^\infty
  \frac{ dx\,x^2}{(x^2 +1)^2 } \left(\frac{m^2+2}{3}\right)^2  \mathfrak{J}_1(m) \nonumber \\
\end{eqnarray}
Here
\begin{equation*}
    \mathfrak{J}_1(m) = \frac{ ( 3 +m^2)T(m) + \frac{1}{2}m^{-2} + 2m^2 L(m) +\frac{1}{2} }{\frac{1}{2}(m^2+1)}
\end{equation*}
Since in our simple model the refractive index $m$ depends only on the two parameters  $x$ and $\omega_P/\omega_0$ , the integral depends only on the ratio $\omega_p/\omega_0 $ of plasma and resonant frequency.
The  {resulting} angular momentum is proportional to the product
of the diamagnetic angular momentum $\hbar \omega_c/\omega_0 $ of  a single dipole, and the number $N = 4\pi a^3 \rho/3 $ of the dipoles.

The extraction of the divergence in $\langle J_{\mathrm{long},z}^{(2)} \rangle$ is harder because of the double integral over
 $k$ and $k'$.  We refer to Appendix~\ref{appC} for details,

\begin{eqnarray}\label{j2final}
   J^{(2)}_{\mathrm{long},z} \approx N\times  \hbar \frac{ \omega_c}{\omega_0}    \frac{1}{2\pi }\int_0^\infty dx\,  \frac{x^2}{(x^2 +1)^2 }
 \left(\frac{m^2+2}{3}\right)^2   \mathfrak{J}_2(m) \nonumber \\
\end{eqnarray}
with $\mathfrak{J}_2(m) = 4 m^2 (m^2-1)/(m^2 + 1)^2 >0$.

\begin{figure}
  \includegraphics[width=0.7\columnwidth, angle=-90]{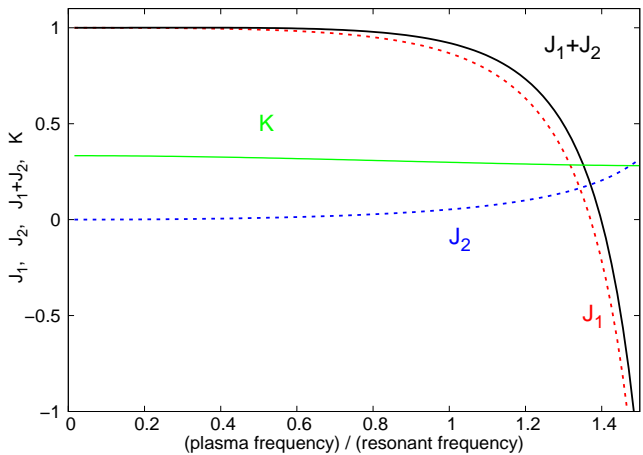}
  \caption{Different coefficients associated with the longitudinal angular momentum of the Mie sphere induced by the  {electromagnetic quantum fluctuations} at $T=0$
  as a function of $\omega_p/\omega_0$. The dashed curves $J_1$ and $J_2$ denote the angular
  momentum given by Eqs.~(\ref{jcanfinal}) and (\ref{j2final}), the solid black curve is the total
  longitudinal angular momentum, all expressed in units of the normal Einstein-De Haas effect $N \hbar \omega_c/\omega_0$.
  {Because electromagnetic modes do not carry magnetization, this curve can be interpreted as the
  inverse Land\'{e} factor of the Mie sphere.}
  The curve labeled $K$ denotes the integral in Eq.~(\ref{Kloc2}) for the curling Poynting vector
  inside the Mie sphere, and is approximately equal to $1/3$ for all $\omega_p/\omega_0$.  {Since total angular momentum is conserved, the angular momentum achieved by the sphere is opposite to
  the electromagnetic angular momentum shown here.}  }
\label{figang}
\end{figure}

The calculations are summarized in Figure~\ref{figang}, where $J^{(1)}_{\mathrm{long},z}$ and $J^{(2)}_{\mathrm{long},z}$ are shown as a function of
$\omega_p/\omega_0$. Clearly $J^{(1)}_{\mathrm{long},z}$ dominates throughout.
For $\omega_p/\omega_0 \ll 1$, {the index of refraction $m( ix)$  is close to unity for all values for $x$ and we recover } $J_{\mathrm{long},z}   = N \times \hbar \omega_c/\omega_0$,
recognized as the   Einstein-De Haas effect  {stemming from the  angular momentum of} $N$ independent diamagnetic dipoles.
{The multiple scattering inside the Mie sphere brings ``extra" electromagnetic angular momentum to the Einstein de Haas effect though with opposite sign compared
to the longitudinal  angular momentum created by the single diamagnetic oscillators. This angular momentum does not carry a magnetic moment, which means that the Land\'{e} factor
of the sphere as a whole, when
defined by the relation $\mathbf{m}= g (e/2mc_0) \mathbf{L}$, is no longer equal to the orbital value $g=1$ (apparently free from QED corrections), but larger.
The opposite sign is not easy to understand heuristically, but this extra angular momentum
is clearly no longer negligible
 when  $\omega_p/\omega_0 \approx 1$.
For larger values, the causality threshold $\omega_p/\omega_0 < \sqrt{3}$ approaches, and the outcome is somewhat speculative in view of the possible invalidity of the Lorenz-Lorentz model.
The Figure shows that the total angular momentum may even change sign at $\omega_p/\omega_0 = 1.4 $. }

\section{A Poynting vortex}
A finite electromagnetic angular momentum along the external magnetic field $\mathbf{B}_0$  implies the existence of an electromagnetic  momentum density $\bar{\mathbf{E}}(\mathbf{r}) \times
\mathbf{B}(\mathbf{r})/4\pi c_0$  circulating around  the external magnetic field $\mathbf{B}_0$. According to
 Eq.~(\ref{noPoyn}), such a curl does neither move optical energy  around, nor does it possess a net momentum.
In the previous section we have seen that
a  part of the angular momentum resides in the bulk, because   {magneto-optical activity}
induces  a  coupling between longitudinal and transverse electric modes.

Let us first address this issue for the  harmonic, diamagnetic oscillator  described by Eq.~(\ref{alpha1}) interacting
with electromagnetic quantum  {fluctuations } at $T=0$.
At any distance from the dipole we can calculate the
expectation value of $\mathbf{K}(\mathbf{r})= \bar{\mathbf{E}}(\mathbf{r}) \times
\mathbf{B}(\mathbf{r})/4\pi c_0$ using the fluctuation-dissipation theorem~(\ref{fd}). We
find a Casimir-Polder type formula,
\begin{equation*}
    \mathbf{K}(\mathbf{r})= - \frac{\hat{\mathbf{r}} \times\hat{\mathbf{ B}}_0}{(4\pi r)^2} \frac{4\hbar \omega_c  }{\pi c_0^3}
    \mathrm{Im}   \int_0^\infty d\omega  \frac{ \omega^5r_e }{(\omega_0^2 - \omega^2 -i0)^2} F\left(\frac{\omega r}{c_0} \right)
\end{equation*}
with $F(x) = \exp(2ix) (1-ix)^2/x^3$.
In the  {non-retarded  regime }  $ r \ll c_0/\omega_0 $ is $F= 1/x^3$ so that
\begin{equation}\label{KvdW}
   \mathbf{K}(\mathbf{r})=  - \hbar \frac{\omega_c}{\omega_0} \frac{r_e}{(4\pi)^2 r^5}  \hat{\mathbf{r}} \times\hat{\mathbf{ B}}_0
\end{equation}
We see that $\mathbf{K}$ is constrained to the dipole and decays rapidly with distance.
The total angular momentum between $r$ and $r+dr$ is $dJ_z \sim +\hbar(\omega_c/\omega_0)   r_e dr/ 4\pi r^2$. The angular momentum
resides at length scales as small as $r \sim  r_e/4\pi $,
consistent with Eq.~(\ref{reguk0}). This conclusion is qualitatively consistent with the quantum mechanical picture of longitudinal
angular momentum being confined  to the charge carriers, as discussed in section I.

We next calculate the momentum density $\mathbf{K}(\mathbf{r})$ inside the Mie sphere, but stay away from the  {discontinuous boundary where angular momentum resides as well}.
Spherical symmetry imposes that $\mathbf{K}(\mathbf{r}) = K(r)
 {\mathbf{r}} \times\hat{\mathbf{ B}}_0$.
The derivation of $K(r)$ reduces to a local version of the total angular momentum derived in the previous section. Since we know longitudinal angular momentum to dominate,
the local version of $J^{(1)}_{\mathrm{long},z}$  {given by Eq}.~(\ref{jcanfinal}) is,

\begin{eqnarray*}
-\frac{2}{3} r K(r)   &=& \frac{2kT}{c_0}\sum_{n=1}^\infty \sum_{\kappa ,\nu =e/m}  \frac{ W(is_n) }{s_n^2/c_0^2 +k^2} \frac{1}{3m^2}    \times  \\
&&  \epsilon_{inl}\epsilon_{imj}   \int \frac{d\hat{\mathbf{r}}}{4\pi } {\bar{E}}^n_{\nu \kappa }(\mathbf{r}) r_l \partial_m {E}^j_{\nu \kappa} (\mathbf{r})
\end{eqnarray*}
or
\begin{eqnarray}\label{Kloc}
K(r)&=& \frac{1}{r} \frac{kT}{c_0  }\sum_{n=1}^\infty\sum_{\kappa ,\nu =e/m} \frac{ W(is_n) }{s_n^2/c_0^2 +k^2} \frac{1}{m^2} \times \nonumber \\
&& \ \  \int \frac{d\hat{\mathbf{r}}}{4\pi }   \left(|\mathbf{E}^T_{\nu\kappa}|^2 + r\partial_r \frac{|\mathbf{E}^T_{\nu\kappa}|^2 - |E^L_{\nu\kappa}|^2 }{2}
\right)
\end{eqnarray}
in terms of the electric field eigenfunctions at $\mathbf{r}$ along  {(L)} and transverse  {(T)} to the vector $\mathbf{r}$.
This expression suffers from the same divergence in the $k-$integral as  {was seen} in the previous section, but here at fixed $r$. The same regularization, using again
{the sums over $J$ derived
in Appendix~\ref{appB}},  gives
\begin{eqnarray*}
 \int \frac{d\hat{\mathbf{r}}}{4\pi }  \sum_{\kappa} \frac{ |\mathbf{E}^T_{m\kappa}(\mathbf{r})|^2 }{s^2/c_0^2 +k^2}\rightarrow  \frac{2}{m^2+1}  \frac{3 }{2r_e}     \\
  \int \frac{d\hat{\mathbf{r}}}{4\pi }   \sum_{\kappa} \frac{|\mathbf{E}^{T(L)}_{e\kappa}(\mathbf{r})|^2 }{s^2/c_0^2 +k^2}\rightarrow  \frac{1(2)m^2}{3}\frac{2}{m^2+1}  \frac{3 }{2r_e}
\end{eqnarray*}
independent of $r$. Thus, inside the Mie sphere and for $T=0$,
\begin{eqnarray}\label{Kloc2}
 && \bm{K}(\mathbf{r})=  \frac{\hat{\mathbf{r}}\times\hat{\mathbf{ B}}_0}{r} \times \frac{3}{2} \rho \hbar \frac{\omega_c}{\omega_0} \times \nonumber \\
  & &  \frac{1}{\pi}
\int_0^\infty  \frac{dx \, x^2}{(x^2+1)^2}   \left(\frac{m^2+2}{3}\right)^2   \frac{2(1+\frac{1}{3}m^2)}{m^2(m^2+1)}
\end{eqnarray}
 The longitudinal angular momentum inside the Mie sphere emerges as a vortex of the  Poynting vector with a
radial profile  $1/r$ from the origin. The integral in  Eq.~(\ref{Kloc2}) is shown in Figure~\ref{figang} and is roughly equal to
$\frac{1}{3}$ for all
$\omega_p/\omega_0$.  {The circulating Poynting vector still diverges in the center of the sphere, but unlike for the single dipole, this singularity poses no problem for
the angular momentum, which is here
homogeneously distributed inside the sphere.}

\section{Discussion, Conclusions and Outlook}

In this work we have considered  {a Lifshitz theory for} the angular momentum of  electromagnetic quantum fluctuations
 interacting with a macroscopic dielectric sphere { whose dielectric function is described by a simple model of electric dipoles with a single resonance}. The angular momentum is induced by {magneto-optical activity}  and proportional to the applied
 magnetic field  {analogous to the traditional  Einstein-de Haas effect, where rotation of an object is induced by its magnetization}.

 We have used the simplest possible model to describe the {magneto-optical activity}  microscopically. In this diamagnetic model the contributions from
 electromagnetic  spin and orbital angular momentum are { seen to be} entirely negligible.
 The underlying atomic diamagnetic angular momentum
 is small - typically of order $\hbar \omega_c/\omega_0$ with $\omega_c$ the cyclotron frequency and $\omega_0$ the resonant frequency of the atoms.
  {The quantum fluctuations excite the electromagnetic modes of the dielectric sphere whose angular momentum  modifies the Einstein de Haas effect.
  More precisely, they lower total angular momentum, but do not affect the diamagnetization - because no charge -
  and therefore increase the Land\'{e} factor of the Mie sphere as a whole. In our theory a relation emerges
 between the Einstein-De Haas effect, the magneto-optical constants and the complete set of eigenfunctions of the Mie sphere over the entire spectral range.
}

To understand the role of  {electromagnetic quantum fluctuations},  the ``necessity of the quantum vacuum"  put forward by Milonni \cite{milonni} is an interesting and relevant
notion since it links the intrinsic diamagnetism of the dipoles itself to the electromagnetic quantum
 fluctuations as well. This notion  brings the Einstein-De Haas effect { for one atom and for the entire dielectric sphere conveniently} on the same footing and provides an
  interesting opportunity to test  the remark by Milonni experimentally.
The major postulate of this work is also consistent with this  view point:  The consistent regularization of a
wave number integral on scales as large as  the inverse electron radius is justified because  macroscopic
longitudinal angular momentum should  still be associated with the microscopic charge carriers.
The explicit temperature dependence of quantum  {fluctuations}
comes in when $kT \sim \hbar \omega_0$. Our formalism acknowledges the temperature dependence, but as long as the atomic resonance $\omega_0$ is
located in the visible regime of the  electromagnetic spectrum, temperature dependence is negligible, as usual in diamagnetism.

{The angular momentum originates from an electromagnetic Poynting vector created by the quantum vacuum fluctuations that circulates around the magnetic field.
It has a singular contribution that only lives at the perfectly discontinuous surface of the sphere.}    Inside the sphere  {electromagnetic} angular momentum emerges as a vortex
in the electromagnetic
 Poynting vector that decays as $1/r$ from the center of the sphere.

 A future challenge is to  {extend} this approach to paramagnetic contributions to {magneto-optical activity} ,  involving
 typical spins of order $\hbar$ and significant dependence on temperature \cite{barron}.  {An important question is if the Einstein-De Haas effect can in general
 be understood as  electromagnetic quantum fluctuations interacting
 with microscopic magneto-dichroism, like for  the special case of diamagnetism discussed here. }

 The author would like to thank Geert Rikken for useful advice.

\appendix

\section{regularization of divergence}\label{appA}

The Vector Helmholtz equation~(\ref{VHH})  for a single electric dipole  can be written as
$p^2\bm{\Delta}_p + \mathbf{V}(\mathbf{r})$, with matter-light interaction
$V(\mathbf{r}) = -\delta(\mathbf{r}) \bm{\alpha}(0) \times \omega^2/c_0^2 $ if we { assume} the electric dipole to be a point-like particle.
In the limit $\omega \rightarrow 0$ the Born expansion applies and $ \mathbf{t}(\omega)$.

\begin{eqnarray*}
    \bm{\alpha}(\omega)  &=& \frac{r_e c_0^2}{\omega_0^2 } \left( 1  + \frac{\omega^2}{\omega_0^2} + { i A \omega } + \cdots \right) \\
   =  &&   \bm{\alpha}(0)   -  \frac{\omega^2}{c_0^2} \bm{\alpha}(0)  \cdot \mathbf{G}_0(\mathbf{r}=0,\omega))\cdot
     \bm{\alpha}(0) + \cdots
\end{eqnarray*}
The first line follows from the expansion of Eq.~(\ref{classdipole}), with $r_e = e^2/mc_0^2$ the classical electron radius. The real part of
$\mathbf{G}_0(\mathbf{r}=0,\omega))$ in the second line diverges.
The longitudinal divergence generates a frequency-independent term in $\bm{\alpha}(\omega)$ and is physically  due to a local field correction to the static polarizability
$\bm{\alpha}(0)$, { of the order of the physical volume of the dipole.} The transverse divergence in $\mathbf{G}_0(\mathbf{0})$ is independent on frequency and negative. { Upon comparing}, we identify
\begin{equation}\label{reguk0}
   \mathrm{Re}\, \mathbf{G}_T(\mathbf{r}=0) = - \frac{2}{3}  \sum_\mathbf{k}   \frac{1}{k^2}  \rightarrow -\frac{1}{r_e}
\end{equation}
The imaginary part does not diverge and $\mathrm{Im}\, \mathbf{G}_0 (\mathbf{r}=0)= -i\omega/6\pi c_0$. {Near the resonant frequency} this identifies
$A= 6\pi c_0/r_e \omega_0^2$ \cite{jackson}.


\section{Sums of spherical Bessel functions} \label{appB}

In this section we derive different sums over magnetic quantum numbers of spherical Bessel function needed to calculate the angular momentum of electromagnetic quantum fluctuations.
We start with the well-know multipole expansion of a plane wave in terms of spherical harmonics \cite{merz}

\begin{equation}\label{plwave}
    \exp(i\mathbf{k}_0\cdot \mathbf{r}) = 4\pi \sum_{JM} i^J j_J(k_0r) Y_J^M(\hat{\mathbf{r}}) \bar{Y}_J^M(\hat{\mathbf{k}}_0)
\end{equation}
The retarded scalar Helmholtz Green function of free space is given by
\begin{equation*}
    \frac{e^{ik|\mathbf{r}-\mathbf{r}'|}}{-4\pi |\mathbf{r}-\mathbf{r}'|} =
    \int\frac{ d^3\mathbf{ k}_0}{(2\pi)^3}  \frac{\exp(i\mathbf{k}_0\cdot \mathbf{r} -i\mathbf{k}_0\cdot \mathbf{r}') }{k^2 -k_0^2 + i0}
\end{equation*}
Inserting Eq.~(\ref{plwave}) and using the closure relation of spherical harmonics
\begin{equation}\label{closure}
    \int d\hat{\mathbf{k}}_0  \, Y_J^M(\hat{\mathbf{k}}_0)Y_{J'}^{M'}(\hat{\mathbf{k}}_0) = \delta_{JJ'}\delta_{MM'}
\end{equation}
the Green function can be expressed as,
\begin{eqnarray}\label{GB}
    \frac{e^{ik|\mathbf{r}-\mathbf{r}'|}}{-4\pi |\mathbf{r}-\mathbf{r}'|} &=& \frac{(4\pi)^2}{(2\pi)^3} \sum_{J=0}^\infty \int_0^\infty dk_0  \frac{k_0^2 j_J(k_0r) j_J(k_0r')}{k^2-k_0^2+i0}\nonumber \\
    &&\ \ \ \ \times \sum_{M=-J}^J Y_J^M(\hat{\mathbf{r}}) \bar{Y}_J^M(\hat{\mathbf{r}}')
\end{eqnarray}
The imaginary part of this identity is equivalent to
\begin{equation}\label{ImGB}
  \mathrm{  sinc } \, k|\mathbf{r}-\mathbf{r}'| = \sum_{J=0}^\infty  (2J+1) j_J(kr) j_J(kr') P_J(\hat{\mathbf{r}}\cdot \hat{\mathbf{r}}')
\end{equation}
with $P_J(\cos\theta)$ the   Legendre polynomial of order $J$. For $\mathbf{r}=\mathbf{r}'$ this reduces to the sum identity
\begin{equation}\label{sum1}
    \sum_{J=0}^\infty  (2J+1) j_J^2(x) =1
\end{equation}
A second sum is obtained by acting on Eq.~(\ref{ImGB}) the  operator
\begin{equation*}
    \mathbf{J}^2 = - \frac{1}{\sin\theta} \frac{d}{d\theta} \sin \theta \frac{d}{d\theta} - \frac{1}{\sin^2\theta} \frac{d^2}{d\phi^2},
\end{equation*}
proportional to the square of the angular momentum in spherical coordinates. Since the Legendre polynomials are eigenfunctions of this operator with eigenvalue $J(J+1)$,  it follows
upon putting  $\mathbf{r}=\mathbf{r}'$ afterwards,
\begin{equation}\label{sum2}
    \sum_{J=0}^\infty  J(J+1)(2J+1) j_J^2(x) = \frac{2x^2}{3 }
\end{equation}
Finally, we can choose the vectors $\mathbf{r}$ and $\mathbf{r}'$ parallel, perform the action $(\partial_r r) (\partial_r' r') $ on Eq.~(\ref{ImGB}) and set  ${r}={r}'$.
This produces the sum,
\begin{equation}\label{sum3}
    \sum_{J=1}^\infty  (2J+1) [(x j_J)']^2 = \sin^2 x + \frac{x^2}{3 }
\end{equation}
We have split off the $J=0$ term because it is absent in electromagnetism.

\section{Bulk-induced longitudinal angular momentum at surface}\label{appC}

We provide an approximate evaluation of the contribution~(\ref{J2}) to the longitudinal angular momentum. The longitudinal eigenfunctions inside the sphere obey
Eq.~(\ref{long}). With normal displacement $\varepsilon E^L$ and potential $\Phi$ continuous at the boundary $r=a$,
and with proper normalization of $\Phi$ far outside the sphere
\cite{merz}, the longitudinal modes inside the sphere are
\begin{eqnarray*}
\bm{E}_{\ell kJM}(\mathbf{r}) &=& \frac{4\pi i^J }{(2\pi)^{3/2}}A^\ell_{Jk} \frac{k }{m}\left[  j_J(kr/m)' \hat{\mathbf{N}}_{JM}(\hat{\mathbf{r}}) \right. \nonumber \\
&& \ \ \  + \left.\sqrt{J(J+1)} \frac{ j_J(kr/m) }{kr/m} \hat{\mathbf{Z}}_{JM}(\hat{\mathbf{r}}) \right]
\end{eqnarray*}
By inserting the eigenfunction into the expression~(\ref{J2}) for the longitudinal angular momentum $ J_{\mathrm{long},z}^{(2)}$  and by carrying out
the angular integrations over $\mathbf{\hat{r}}$, using general properties of the vector spherical harmonics \cite{cohen,lacoste}, we get the still rigorous expression,
\begin{eqnarray*}
 &&  J^{(2)}_{\mathrm{long},z} = -\frac{\hbar a^2}{3\pi c_0} \frac{(4\pi)^4 }{(2\pi)^6} \int_0^\infty ds\, W(is) \frac{m^2-1}{m^2}\sum_{kk'} \sum_J   \\
&&  \ \ \times (2J+1) |A^e_{kJ}|^2 |A^\ell_{kJ}|^2  \frac{k'^2}{s^2/c_0^2 + k^2} (yj_J)'(mka) j'_J(k'a/m) \\
&&  \times \int_0^\infty  dr\, r \theta_a \left\{ \frac{j_J }{x} (yj_J)' + (J(J+1) \frac{j_J }{x}j_J(y) + j_J'(x)(yj_J)' \right\}
\end{eqnarray*}
with $x = k'r/m$, $y= mkr$ and $\sum_k = \int_0^\infty dk$ . This expression diverges for large wave numbers but since two $k$-integrals exist the extraction of the divergence is more complicated.
Our approximation will be that we consider this expression for\emph{ both }large $k$ and $k'$ in which case
\begin{equation*}
         \left|A^\ell_{Jk}\right|^2 \rightarrow  \frac{2m^2 }{m^2 +1 + (m^2-1) \cos(2ka/m -J\pi )}
\end{equation*}
We ignore all functions that oscillate with $k$ and that produce finite corrections for the longitudinal momentum, proportional to the surface.
To perform  the integral involving $k'$, we use
\begin{eqnarray*}
  \sum_k k^2 j_J'(k a) \frac{j_J(kr)}{kr} &=&-\frac{\pi}{2} \frac{J+1}{2J+1} \frac{r^{J-1}}{a^{J+2}}  \ \ (r \leq a) \\
   \sum_k k^2 j_J'(k a) {j'_J(kr)}&=& \frac{\pi}{2}\left( \frac{\delta(r-a)}{a^2} - \frac{J(J+1)}{2J+1}\frac{r_<^{J-1}}{r_>^{J+2}}  \right)
\end{eqnarray*}
with neglect of terms of measure $0$ at $r=a$. The result is,
\begin{eqnarray*}
   J^{(2)}_{\mathrm{long},z} \approx && -\frac{\hbar a }{6 c_0} \frac{(4\pi)^4 }{(2\pi)^6} \int_0^\infty ds\, W(is)\frac{ 4 m^2 (m^2-1)}{(m^2 + 1)^2} \times  \\
&&   \left( - \sum_{J=1}^\infty (J+1)^2 \partial_a \left[a I_J(r,a)\right]_{r=a}  \right. \\
&& \ \ \  \left. + \int dr \theta_a \delta(r-a) \sum_{J=1}^\infty (2J+1) (\partial_a a)( \partial_r r) I_J(r,a)  \right)
\end{eqnarray*}
with the definition
\begin{equation*}
    I_J(r,a) \equiv \sum_k  \frac{j_J(kr) j_J(ka) }{k^2}
\end{equation*}
This function can be expressed using the  hypergeometric function $F(J-\frac{1}{2},J+1,2J+2,z)$ \cite{GR684}. From this we can show that  $\partial_a [aI_J(r,a)]_{r=a}\sim a/J^4$ for large $J$.
Hence the first term in  $J^{(2)}_{\mathrm{long},z}$ is finite and  {generates an angular momentum }of order $\hbar (\omega_c/\omega_0) \times \rho r_e a^2$, which is
negligible with respect to what was found in Eq.~(\ref{jcanfinal}).
Using the sum Eq.~(\ref{ImGB}) in Appendix~\ref{appB}, the second term follows from
\begin{eqnarray*}
  &&  \sum_{J=1}^\infty (2J+1) (\partial_a a)( \partial_r r) I_J(r,a)  = \\
    && \ \ \ \  \sum_k \frac{1}{k^2} (\partial_a a)(\partial_r r) \left(\frac{ \sin k(a-r)}{k(a-r)} - \frac{\sin ka}{ka}  \frac{\sin kr}{kr}  \right) \\
    && \ \ \  =     \sum_k \left( \frac{\sin^2 ka }{k^2} - \frac{a^2}{3} \right) \ \ \mathrm{for}\  r=a
\end{eqnarray*}
Regularizing  as in Appendix~\ref{appA} $4\pi /(2\pi)^3 \int dk = 3 /2r_e$ and from Eq.~(\ref{truc}) $\int dr \theta_a \delta(r-a) = 1/2$, we find
 \begin{eqnarray}\label{j2finalapp}
   J^{(2)}_{\mathrm{long},z} \approx  \frac{\hbar a^3 }{3 c_0 r_e} \int_0^\infty ds\, W(is)\frac{ 4 m^2  (m^2-1)}{(m^2 + 1)^2}
\end{eqnarray}


\begin{thebibliography}{4}
\bibitem{notegz} For the quarter circle $\omega = R \exp(i\phi)$  in the Cauchy contour
involving Eq.~(\ref{jz}),  with $0\leq \phi \leq \pi/2$  and $R\rightarrow \infty$, we have
$\omega^2 W(\omega) \rightarrow \mathrm{real\, constant}$; For large $R$ we can work in the Born approximation and perform the trace in
Fourier space. For the transverse angular momentum  this produces a factor $i\omega/c_0$ and we end up with the  imaginary
part of a real-valued number. The longitudinal angular momentum of the quarter circle vanishes as $1/R$.
\bibitem{notU} Since $\bm{\nabla} \cdot\mathbf{ K}= 0$ everywhere for the quantum vacuum expectation value of the momentum density $\mathbf{K}$, it can be written as
$\mathbf{K}= \frac{1}{2}  \bm{\nabla} \times \mathbf{j}({r})$. Then $\mathbf{J}(r<a) = -\mathbf{j}(a) V(a) + \int_{V(a)} d^3 \mathbf{r } \mathbf{j}(r)$, that is
the sum of a surface term and a bulk term. However,
$\mathbf{j}(r)$ is only determined up to a constant
vector.
\end{thebibliography}
\end{document}